\documentclass[aps,prl,twocolumn,nofootinbib,superscriptaddress]{revtex4}
\usepackage{amsmath,bm}
\usepackage{graphicx}
\usepackage{natbib}

\newcommand{\var}[1]{\mathrm{Var}\left( #1 \right)}
\newcommand{\M}[1]{\left\langle  #1 \right\rangle}
\newcommand{\lr}[1]{\left(  #1 \right)}

\usepackage{amsmath,amsfonts,amssymb,graphics,graphicx,epsfig,color,times,bbm}

\newcommand{\tr}[1]{{\rm tr}\left[#1\right]}

\begin{document}

\title{Quantum memory for entangled two-mode squeezed states}

\author{K. Jensen}
\affiliation{QUANTOP, Danish National Research Foundation Center for Quantum Optics, Niels Bohr Institute, University of Copenhagen, DK 2100, Denmark}

\author{W. Wasilewski}
\affiliation{QUANTOP, Danish National Research Foundation Center for Quantum Optics, Niels Bohr Institute, University of Copenhagen, DK 2100, Denmark}

\author{H. Krauter}
\affiliation{QUANTOP, Danish National Research Foundation Center for Quantum Optics, Niels Bohr Institute, University of Copenhagen, DK 2100, Denmark}

\author{T. Fernholz}
\affiliation{QUANTOP, Danish National Research Foundation Center for Quantum Optics, Niels Bohr Institute, University of Copenhagen, DK 2100, Denmark}

\author{B. M. Nielsen}
\affiliation{QUANTOP, Danish National Research Foundation Center for Quantum Optics, Niels Bohr Institute, University of Copenhagen, DK 2100, Denmark}

\author{A. Serafini}
\affiliation{University College London, Department of Physics $\&$ Astronomy, Gower Street, London WC1E 6BT}

\author{M. Owari}
\affiliation{Institut f{\"u}r Theoretische Physik, Universit{\"a}t Ulm,
Albert-Einstein Allee 11, D-89069 Ulm, Germany}

\author{M. B. Plenio}
\affiliation{Institut f{\"u}r Theoretische Physik, Universit{\"a}t Ulm,
Albert-Einstein Allee 11, D-89069 Ulm, Germany}

\author{ M. M. Wolf}
\affiliation{QUANTOP, Danish National Research Foundation Center for Quantum Optics, Niels Bohr Institute, University of Copenhagen, DK 2100, Denmark}

\author{E. S. Polzik}
\affiliation{QUANTOP, Danish National Research Foundation Center for Quantum Optics, Niels Bohr Institute, University of Copenhagen, DK 2100, Denmark}

\maketitle

\textbf{
A quantum memory for light is a key element for the realization of future quantum information networks \cite{Kimble:2008,Lvovsky:2009,Hammerer09}. Requirements for a good quantum memory are (i) versatility (allowing a wide range of inputs) and (ii) true quantum coherence (preserving quantum information). Here we demonstrate such a quantum memory for states possessing Einstein-Podolsky-Rosen (EPR) entanglement. These multi-photon states are two-mode squeezed by $6.0$ dB with a variable orientation of squeezing and displaced by a few vacuum units. This range encompasses typical input alphabets for a continuous variable quantum information protocol. The memory consists of two cells, one for each mode, filled with cesium atoms at room temperature with a memory time of about $1$msec. The preservation of quantum coherence
is rigorously proven by showing that the experimental memory fidelity $0.52\pm0.02$ significantly exceeds the benchmark of $0.45$ for the best possible classical memory for a range of displacements.}

A sufficient condition for a memory to be genuinely quantum can be formulated via the fidelity between memory input and output. If this fidelity exceeds the benchmark determined by the best classical device then it can store and preserve entanglement and hence is a true quantum memory.
Classical benchmark fidelities are difficult to calculate, and until recently they were known only for coherent states \cite{Hammerer:2005} and homogeneously distributed pure states \cite{Massar:1995}. Storage of non-classical correlations between single photons \cite{Kuzmich03,Chaneliere05,Eisaman05} and a dual-path superposition state of a photon \cite{Choi08} have been experimentally demonstrated with the Raman and electro-magnetically induced transparency (EIT) approaches. Alongside these demonstrations for single photons, quantum memory for multi-photon entangled states is an important and challenging ingredient of quantum information networks.

A particular class of multi-photon entangled states, namely EPR-type entangled two-mode squeezed states, plays a fundamental role in quantum information processing with continuous variables (cv) \cite{Ralph:2009, Furusawa:2006}. A quantum memory for entangled cv states is valuable for
iterative continuous variable entanglement distillation \cite{Browne:2003},
continuous variable cluster state quantum computation \cite{Gu:2009,Menicucci:2006}, communication/cryptography protocols involving several rounds \cite{Lamoureux:2005}, and quantum illumination \cite{Lloyd:2008,Tan:2008}.
Temporal delay of EPR entangled states has been demonstrated in \cite{Marino:2009}. Very recently, classical benchmarks for storing a squeezed vacuum state \cite{Adesso08} and displaced squeezed states \cite{Owari09,Calsamiglia:2009} have been derived. EIT-based memory for squeezed vacuum  has been recently reported \cite{Appel08,Honda08}, albeit with the fidelity below the classical benchmark.

Here we report the realization of a quantum memory for a set of displaced two-mode squeezed states with an unconditionally high fidelity that exceeds the classical benchmark \cite{Owari09}.

\begin{figure*}
\centering
\includegraphics[scale=.7]{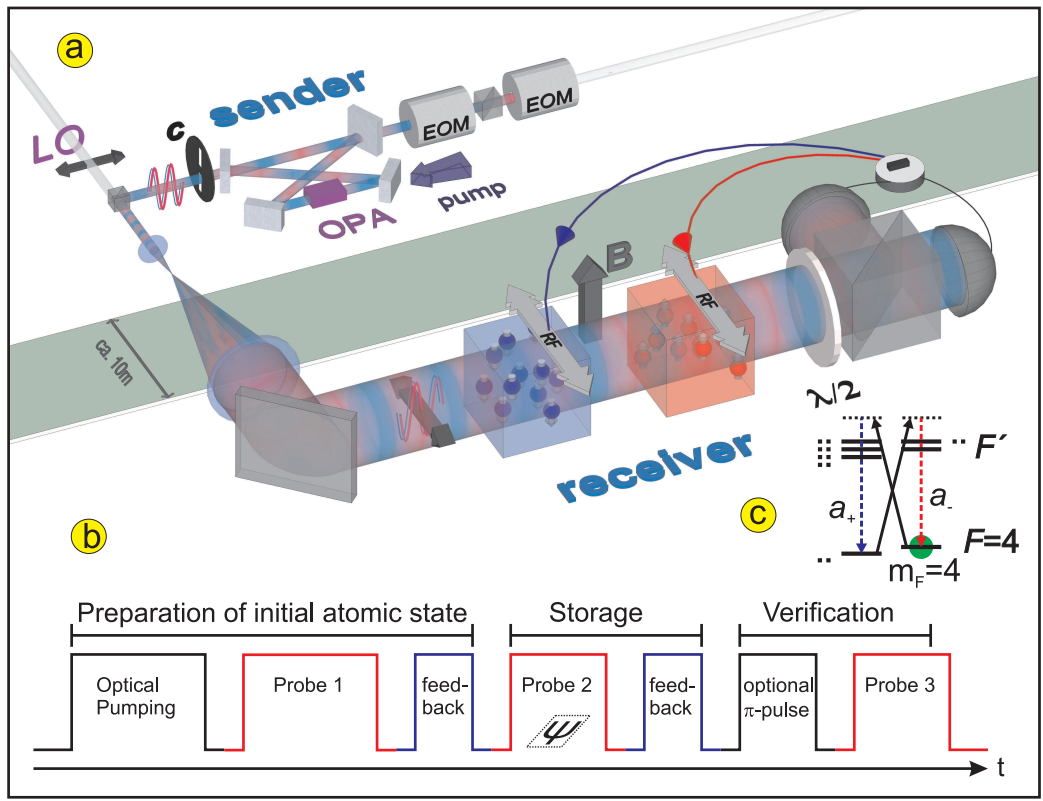}
\caption{\textbf{Setup and pulse sequence.} a. At the sender station two-mode entangled (squeezed) light is generated by the Optical parametric Amplifier (OPA). A variable displacement of the state is achieved by injecting a coherent input into OPA modulated by electro-optical modulators (EOM). The output of the OPA is shaped by a chopper, and combined on a polarizing beamsplitter with the local oscillator (LO) beam, such that the squeezed light is only on during the second probe pulse. A beamshaper and a telescope create an expanded flattop intensity profile. The light is then send to the receiver (memory) consisting of two oppositely oriented ensembles of spin-polarized cesium vapour in paraffin coated cells and a homodyne detector. The detector signal is processed electronically and used as feedback onto the spins obtained via RF magnetic field pulses. b. Pulse sequence for the initiation of the memory, storage, and verification. c. Atomic level structure illustrating interaction of quantum modes with the memory.}
\label{fig:setup}
\end{figure*}

 We store a displaced EPR state of two modes of light $\hat{a}_{+}$ and $\hat{a}_{-}$ with the frequencies $\omega_{\pm}=\omega_0\pm\omega_L$, where $\omega_0$ is the carrier frequency of light. The entanglement condition is $\rm{Var}(\hat{X}_{+}+\hat{X}_{-})+\rm{Var}(\hat{P}_{+}-\hat{P}_{-})<2$ \cite{Duan:2000} where canonical quadrature operators obey $[\hat{X}_\pm,\hat{P}_\pm]=i$.
 For a vacuum state $\rm{Var}(\hat{X}_{\rm{vac}})=\rm{Var}(\hat{P}_{\rm{vac}})=1/2$.
The EPR entanglement of the $\hat{a}_{+}$ and $\hat{a}_{-}$ modes is equivalent to simultaneous squeezing of the $\cos(\omega_{L} t)$ mode $\hat{x}_{Lc}=(\hat{X}_{+}+\hat{X}_{-})/\sqrt{2}$; $\hat{p}_{Lc}=(\hat{P}_{+}+\hat{P}_{-})/\sqrt{2}$ and the corresponding $\sin(\omega_{L} t)$ mode. The alphabet of quantum states is generated by displacing the two-mode squeezed vacuum state by varying values $\langle\hat{x}_{Lc,s}\rangle$ and $\langle\hat{p}_{Lc,s}\rangle$ and by varying the orientation of the squeezed quadrature between $\hat{x}_{L}$ and $\hat{p}_{L}$.
The displaced squeezed states are produced (Fig.\ \ref{fig:setup}a) using an optical parametric amplifier (OPA) \cite{Schori:2002} with the bandwidth of $8.3$MHz and two electro-optical modulators (EOMs)(see the Methods summary for details).

 The two photonic modes are stored in two ensembles of cesium atoms contained in paraffin coated glass cells (Fig.\ \ref{fig:setup}a) with the ground state coherence time around $30$msec \cite{Hammerer09}. $\omega_0$ is blue detuned by $\Delta=855$ MHz from the $F=4\rightarrow F'=5$ of D2 transition (Fig.\ \ref{fig:setup}c). Atoms are placed in a magnetic field which leads to the precession of the ground state spins with the Larmor frequency $\omega_L=2\pi\cdot322$ kHz. This ensures that the atoms efficiently couple to the entangled $\omega_{\pm}=\omega_0 \pm \omega_L$ modes of light.
 The two ensembles $1(2)$ are optically pumped in $F=4, m_{F}=\pm 4$ states, respectively, which leads to the opposite orientation of their macroscopic spin components $J_{x1}=-J_{x2}=J_x$.

  The atomic memory is conveniently described by two sets $c,s$ of non-local, i.e., joint for the two separate memory cells, canonical atomic operators \cite{Julsgaard03}
$x_{Ac} = (J^{rot}_{y1}-J^{rot}_{y2})/\sqrt{2 J_x},p_{Ac} = (J^{rot}_{z1}+J^{rot}_{z2})/\sqrt{2J_x},x_{As} = -(J^{rot}_{z1}-J^{rot}_{z2})/\sqrt{2J_x},p_{As} = (J^{rot}_{y1}+J^{rot}_{y2})/\sqrt{2J_x}$
where the superscript $rot$ denotes spin operators in a frame rotating at $\omega_L$.
 It can be shown that the cosine/sine light mode couples only to the atomic $c/s$ mode, respectively. As a consequence, in the protocol described below the upper(lower) entangled sideband mode of light is stored in the $1(2)$ memory cell, respectively. Since the equations describing the interaction are the same, we omit the indices $c,s$ from now on.

 Light sent from the sender station to the receiver (memory) station consists of quantum $x$-polarized modes and a strong $y$-polarized part which serves as the driving field for interaction with atoms and as the local oscillator (LO) for the subsequent homodyne measurement (Fig.\ \ref{fig:setup}).
The interaction of light and a gas of spin polarized atoms under our experimental conditions can be described by the following input-output equations \cite{Wasilevski09}:
\begin{align}
x_A' =& \sqrt{1-\frac{\kappa^2}{Z^2}}x_A +\kappa p_L,&
p_A' =& \sqrt{1-\frac{\kappa^2}{Z^2}}p_A -\frac{\kappa}{Z^2} x_L,& \nonumber\\
x_L' =& \sqrt{1-\frac{\kappa^2}{Z^2}}x_L +\kappa p_A,&
p_L' =& \sqrt{1-\frac{\kappa^2}{Z^2}}p_L -\frac{\kappa}{Z^2} x_A ,& \label{eq:io}
\end{align}
where the coupling constant $\kappa$ is a function of light intensity, density of atoms and interaction time, and $Z^2=6.4$ is a function of the detuning only. In the limit $\kappa \rightarrow Z$ these equations describe a swap of operators for light and atoms, i.e., a perfect memory followed by squeezing by a factor $Z^2$.
However, in our experiment the swapping time is too long compared to the atomic decoherence time. We therefore follow another approach which takes a much shorter time.

The sequence of operations of the quantum memory protocol is shown in Fig.\ \ref{fig:setup}b. After the light-atoms interaction during the storage part of the protocol the output light operator $x_L'$ is measured by the polarization homodyne detection  and the result is fed back onto the $p_A'$ with a gain $g$. This feedback into both $c$ and $s$ modes is achieved by applying two pulses of the magnetic field at the frequency $\omega_{L}$ to the two cells. %\cite{Julsgaard04}.
The resulting $x_A^{\mathrm{fin}}$ and $p_A^{\mathrm{fin}}$ for the optimized $g$ and $\kappa=1$ can be found from Eq.\ (\ref{eq:io})

\begin{equation}
x_A^{\mathrm{fin}} = \sqrt{1-\frac{1}{Z^2}}x_A + p_L  \ \ \mathrm{and} \ \ p_A^{\mathrm{fin}}=-x_L. \label{eq:mem}
\end{equation}

In the absence of decoherence the operator $x_{L}$ is perfectly mapped on the memory operator $p_A^{\mathrm{fin}}$. The operator $p_{L}$ is stored in $x_A^{\rm{fin}}$ with the correct mean value $\langle p_{L}\rangle = \langle x_A^{\rm{fin}}\rangle$ (since $\langle x_A\rangle=0$) but with an additional noise due to $x_A$.

 The ability to reproduce the correct mean values of the input state in the memory by adjusting $g$ and $\kappa$ is a characteristic feature of our protocol which arguably makes it better suited for storage of multi-photon states compared to, e.g., EIT approaches where such an adjustment is not possible.

 The additional noise of the initial state of atoms $x_A$ is suppressed by the factor $\sqrt{1-\frac{1}{Z^2}}=0.92$ due to the swap interaction. To reduce it further we start the memory protocol with initializing the atomic memory state in a spin-squeezed state (SSS) with a squeezed $x_A$. The SSS with variances $\var{x_A}=0.43(3)$ and $\var{p_A}=1.07(5)$ is generated as in \cite{Julsgaard01} by the sequence shown as the preparation of initial state in Fig.\ \ref{fig:setup}b following optical pumping of atoms into the state with $\var{x_A}=0.55(4)$.

 Before the input state of light undergoes various losses it is a $6$dB squeezed state which we refer to as an "initial pure state". In the photon number representation for the two modes the state is $|\Psi\rangle=0.8|0\rangle_{+}|0\rangle_{-}+0.48|1\rangle_{+}|1\rangle_{-}+ 0.29|2\rangle_{+}|2\rangle_{-}+0.18|3\rangle_{+}|3\rangle_{-}+...$. We ran the memory protocol for 18 different states defined by  $\left\{ \phi=0,90 \ \mathrm{deg}; [\M{x_L} ;\M{p_L}]=[0,3.8,7.6;0,3.8,7.6] \right\}$, where $\phi$ is the phase of the squeezing, and $\M{x_L}$ and $\M{p_L}$ are the displacements of the initial state (see Table \ref{table1}). In the absence of passive (reflection) losses for light and atomic decoherence we find from Eq.\ (\ref{eq:mem}) the expected fidelity of $0.95$, $0.61$ for the states squeezed with the $\phi=0$, $\phi=90$, respectively, with the mean fidelity of $0.78$.

In the experiment light is sent to the receiver's memory station via a transmission channel with the transmission coefficient $\eta_{\rm{loss}}=.80(4)$ (which includes the OPA output coupling efficiency) resulting in the "memory input state" with $\var{x_L\cdot cos( \phi)-p_L\cdot sin(\phi)}=0.20(2)$ and $\var{x_L\cdot sin( \phi)+p_L\cdot cos(\phi)}=1.68(9)$. The state of light is further attenuated by the factor $\eta_{\rm{ent}}=.90(1)$ due to the entrance (reflection) losses at the windows of the memory cells. Between the interaction and detection light experiences losses described by the detection efficiency $\eta_{\rm{det}}=.79(2)$ (see Methods Summary).

Following the storage time of $1$msec we measure the atomic operators with the verifying probe pulse in a coherent state. The mean values and variances of the atomic operators $x_A^{\rm{fin}}$ and $p_A^{\rm{fin}}$ are summarized in Table \ref{table1} (see the Methods summary for calibration of the atomic operators).
From these values and the loss parameters, we can calculate the noise added during the storage process which is not accounted for 
by transmission and entrance losses.
We find that the memory adds $0.47(6)$ to $\rm{Var} (x_A^{\rm{fin}})$ and $0.38(11)$ to $\rm{Var} (p_A^{\rm{fin}})$, whereas for the ideal memory, according to Eq.\ (\ref{eq:mem}), we expect the additional noise to be $0.36(5)$(due to the finite squeezing of the initial atomic operator $x_A$) and $0$, for the two quadratures respectively. This added noise can be due to atomic decoherence, uncanceled noise from the initial anti-squeezed $p_A$ quadrature, and technical noise from the EOMs.

\begin{figure}
\centering
\includegraphics[scale=0.5]{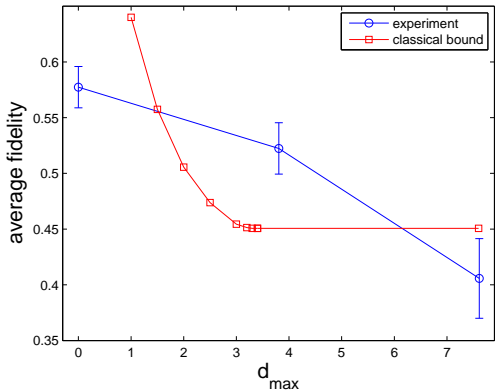}
\caption{\textbf{Fidelities.}
The fidelity $F$ calculated from experimental results is shown as circles connected by lines. The theoretical benchmark values are shown as squares.
The horizontal axis is the size $d_{\rm{max}}$ of the input distribution. Note that one vacuum unit of displacement corresponds to $d_{\rm{max}}=1/\sqrt{2}$.}
\label{fig:fidelities}
\end{figure}

The overlap integrals between the stored states and the initial pure states are given in Table \ref{table1}.
The average fidelities calculated from the overlap values for square input distributions with the size $d_{\rm{max}}=0,3.8 \ \rm{and} \ 7.6$ are plotted in
Fig.\ \ref{fig:fidelities}. The choice of the interaction strength $\kappa=1$ minimizes the added noise but leads to the mismatch between the mean values of the stored atomic state and of the initial pure state of light by the factor $\sqrt{\eta_{\rm{ent}}\cdot\eta_{\rm{loss}}}=0.85$. This mismatch is the reason for the reduction of the experimental fidelity for states with larger displacements.

The experimental fidelity is compared with the best classical memory fidelity which is calculated \cite{Owari09} from the overlap of the initial pure state with the state stored in the classical memory positioned in place of the quantum memory. The input distribution used in the calculations is a square $\left\{ \left|\M{x_L}\right|, \left|\M{p_L}\right|\leq d_{\rm{max}} \right\}$
with all input states with the mean values within the square having equal probability and other states having probability 0. The squeezing of the input states is fixed to the experimental value, and all phases of squeezing are allowed. The upper bound on the classical benchmark fidelity is plotted in Fig.\ \ref{fig:fidelities}.
The benchmark values have been obtained by first truncating
the Hilbert space to a finite photon number and then solving the finite
dimensional optimization employing semi-definite
programming. The truncation of the Hilbert space is treated rigorously
by upper bounding its effect via an error bound that is included in the
fidelity. As a result, around the value $d_{\rm{max}}= 3.5$ the
theoretical calculated fidelity does not decrease further, but remains at
a constant level due to the rapidly increasing truncation error (see the Supplementary Material for the proof of the non-increasing character of the dependence of the benchmark on $d_{\rm{max}}$).

For the input distributions around  $d_{\rm{max}}=3.8$ (a vacuum unit of displacement is $d= 1/\sqrt{2}$), the measured fidelity is higher than the classical bound and thus the memory is a true quantum memory.

Outperforming the classical benchmark means that our memory is capable of preserving EPR-entanglement in case when one of the two entangled modes is stored while the other is left propagating. Using experimentally obtained values of the added noise  we evaluate the performance of our memory for the protocol where the upper sideband $\hat{a}_+$ mode is stored in one of the memory cells whereas the other EPR mode $\hat{a}_-$ is left as a propagating light mode. We find an EPR variance between the stored mode and the propagating mode of $1.52$ ($-1.2$dB below the separability criterion) which corresponds to the lower bound on the entanglement of formation of $\sim1/7$ebit \cite{Giedke:2003} (see Supplementary Information for details of the calculation). Implementing this version of the memory would require spectral filtering of the $\hat{a}_-$ and $\hat{a}_+$ modes which can be accomplished by a narrow band optical cavity.

In conclusion, we have experimentally demonstrated deterministic quantum memory for multi-photon non-classical states. An obvious way to improve the fidelity is to  reduce the reflection losses on the memory cell windows. Other improvements involve increasing the initial atomic spin squeezing and reduction of atomic decoherence which should allow for the storage protocol based on the swap interaction.

\subsection{Methods Summary}

\paragraph{\textbf{Verification.}}
By measuring the $x_L'$ of the verification pulse the statistics of the atomic operator $p_A^{\rm{fin}}$ is obtained. In another series we rotate $x_A^{\rm{fin}}$ into $p_A^{\rm{fin}}$ (and vice versa) using a magnetic $\pi$-pulse and obtain the statistics of $x_A^{\rm{fin}}$ for the same input state of light.
Since we assume Gaussian statistics, the mean values and the variances of $x_A^{\rm{fin}}$ and $p_A^{\rm{fin}}$ are sufficient for a complete description of the atomic state.

\paragraph{\textbf{Calibrations.}}
Before performing the storage, we calibrate the interaction strength $\kappa$ and the feedback gain $g$, such that the mean values of the light state inside the memory (i.e. after the entrance loss) are transferred faithfully.
$\kappa$ is calibrated by creating a mean value $\M{p_L^{\rm{2nd}}}$ in the second probe pulse. The mean is stored in the atomic $x_A'$, which is read out after the magnetic $\pi$-pulse with the third probe pulse. The measured mean of the third pulse is then $\M{x_L'^{\rm{3rd}}}=\kappa^2 \M{p_L^{\rm{2nd}}}$ from which $\kappa^2$is determined.
Using similar methods we can calibrate the electronic feedback gain $g$.

\paragraph{\textbf{Generation of the displaced squeezed input states.}}
The OPA which is pumped by the second harmonic of the master laser and generates the entangled squeezed vacuum states is seeded with a few $\mu$W of the master laser
light
%the $\omega_{0}$ beam
with the carrier frequency $\omega_{0}$ which is amplitude and phase modulated by two electro-optical modulators (EOMs) at a frequency of 322 kHz, thus creating coherent states in the $\pm$ 322 kHz sidebands around $\omega_{0}$ (Fig.\ \ref{fig:setup}). With such a modulated seed, the output of the OPA is a displaced two-mode squeezed state. The output of the OPA is mixed at a polarizing beamsplitter with the strong local oscillator (driving) field from the master laser.

\paragraph{\textbf{Losses.}}
In order to calculate the mean values and variances of the stored state and the input states, we need to know the optical losses. We choose to divide the total losses $\eta_{\mathrm{tot}}$ into three parts, the channel propagation losses $\eta_{\mathrm{loss}}$ from the OPA to the front of the memory cells (including the OPA output efficiency), the entrance loss $\eta_{\mathrm{ent}}$ and the detection losses $\eta_{\mathrm{det}}$, such that $\eta_{\mathrm{tot}}$=$\eta_{\mathrm{loss}} \cdot \eta_{\mathrm{ent}} \cdot \eta_{\mathrm{det}}$ (all the $\eta$'s are intensity transmission coefficients).
From the measurement of the quadratures of the squeezed light (with variances $0.29(1)$ and $1.34(6)$, we find the total losses $\eta_{\rm{tot}} =0.567 (35)$.
We measure the transmission through the cells of 0.817(20), the transmission through the optics after the cells of 0.889(10) and estimate the efficiency of the photodiodes to be 0.98(1). Assigning one half of the losses through the cells to the entrance losses and another half to the detection losses we find $\eta_{\rm{ent}}=\sqrt{0.817}=0.90(1)$,
$\eta_{\rm{det}}=\sqrt{0.817}\cdot0.889\cdot 0.98=0.79(2)$ and
$\eta_{\rm{loss}}=\eta_{\rm{tot}}/\left( \eta_{\rm{ent}} \eta_{\rm{det}} \right)=0.80(4)$.

\begin{table}
	\centering
		\begin{tabular}{|c|c|c|r|r|c|c|c|}
		\hline
\multicolumn{3}{|c|}{Input states} & \multicolumn{4}{|c|}{Stored states} & overlap \\ \hline

		$\M{x_L}$ & $\M{p_L}$ & $\phi$ & $\  \M{p_A^{\mathrm{fin}}} $ & $\ \M{x_A^{\mathrm{fin}}}$ & $\var{p_A^{\mathrm{fin}}}$ & $\var{x_A^{\mathrm{fin}}}$ & $F$ \\ \hline
   0.0 &   0.0 &     0 & -0.06 &  0.25 &  0.52(2) &  1.99(3) &  0.62\\
   0.0 &   3.8 &       & -0.06 &  3.19 &   				&          &  0.60\\
   3.8 &   0.0 &       & -3.47 & -0.42 &   				&  				 &  0.57\\
   3.8 &   3.8 &       & -3.39 &  2.89 &   				&          &  0.49\\ \hline

   0.0 &   0.0 &    90 & -0.07 &  0.06 &  1.95(6) &  0.73(1) &  0.55\\
   0.0 &   3.8 &       & -0.06 &  3.14 &   				&          &  0.42\\
   3.8 &   0.0 &       & -3.22 &  0.48 &   				&          &  0.46\\
   3.8 &   3.8 &       & -3.21 &  3.59 &  				&          &  0.50\\ \hline \hline

   0.0 &   7.6 &     0 & -0.03 &  6.30 & 0.55(2) &  2.01(4) &  0.49\\
   7.6 &   0.0 &       & -6.83 & -0.46 &         &          &  0.37\\
   3.8 &   7.6 &       & -3.20 &  6.07 &         &          &  0.35\\
   7.6 &   3.8 &       & -6.54 &  2.80 &         &          &  0.22\\
   7.6 &   7.6 &       & -6.40 &  6.03 &         &          &  0.15\\ \hline

   0.0 &   7.6 &    90 & -0.08 &  6.24 &  2.12(8) &  0.78(3) &  0.18\\
   7.6 &   0.0 &       & -6.37 &  0.59 &          &          &  0.35\\
   3.8 &   7.6 &       & -3.13 &  6.75 &          &  				 &  0.32\\
   7.6 &   3.8 &       & -6.38 &  3.79 &     			& 				 &  0.43\\
   7.6 &   7.6 &       & -6.36 &  6.72 &  				& 				 &  0.27\\ \hline
		\end{tabular}
		
\caption{The three first columns display the mean displacements and the squeezing phase ($\phi=0$ corresponds to $x_L$ being squeezed) of the initial pure light states. The next four columns display the mean values and variances of the atomic states after the storage. 
The last column displays the overlap between the initial pure light states and the stored atomic states. Vacuum state variances are $0.5$.
The uncertainties on the variances are calculated as the standard deviation of the variances within each subgroup of the input states.}\label{table1}
\end{table}

\bibliographystyle{apsrev}
\bibliography{v4}

\section{Acknowledgments}
This work was supported by EU projects QESSENSE, COMPAS and COQUIT.

\section{Author Contributions}
Experimental group: K. J., W. W., H. K., T. F., B. M. N. and E. S. P. Calculation of the classical benchmark: M. O., M. B. P., A. S. and M. M. W.

%%%%%%%%%%%%%%% Supplementary Information %%%%%%%%%%%%%%%%%%%%%%%%%

\clearpage

\section{Supplementary Information}
\subsection{Memory added noise}

As discussed in the main text, in the experiment the final atomic states acquire extra noise from losses and decoherence not included in the basic theory.
This extra noise is accounted for in two steps. In the first step the initial pure light state propagates  through the transmission and memory entrance losses $\eta_{\rm{loss}} \eta_{\rm{ent}}$ (modelled by a beamsplitter) which add vacuum noise to the state of light. In the second step extra noise is added in the process of performing a unity-gain storage of the light state. % (with canonical operators denoted $x_L^{\rm{in}}$ and $p_L^{\rm{in}}$).
We model this extra added noise with two operators $S_x$ and $S_p$ with zero mean values. The final atomic state is now (see also Eq.\ (2) in the main text)
\begin{align}
x_A^{\mathrm{fin}} =&  \sqrt{1-\frac{1}{Z^2}}x_A + G p_L^{\rm{pure}} +\sqrt{1-G^2} p_L^{\rm{vac}}  + S_x,& \nonumber \\
p_A^{\mathrm{fin}}=& -G x_L^{\rm{pure}} -\sqrt{1-G^2}x_L^{\rm{vac}} +S_p \label{eq:mem2},
\end{align}
where we defined the gain of the memory $G=\sqrt{\eta_{\rm{loss}} \eta_{\rm{ent}}}=0.85$, which equals the ratio of the mean values of the stored atomic state and the intial pure state (this gain should not be mistaken with the electronic feedback gain).
$x_{L}^{\rm{vac}}$ and $p_{L}^{\rm{vac}}$ are vacuum operators with the zero mean and the variance 1/2.
Taking into account that $Z^2=6.4$ and the initial atomic noise is $\var{x_A}=0.43$ we can calculate the expected variances of the final atomic state from the above equation. 
\begin{align}
\var{x_A^{\mathrm{fin}}} =&  \lr{1-\frac{1}{Z^2}} \var{x_A} +G^2 \var{p_L^{\rm{pure}}} & \nonumber \\
                          &+\lr{1-G^2} \frac{1}{2}  + \var{S_x}, & \nonumber \\
\var{p_A^{\mathrm{fin}}}=& G^2 \var{x_L^{\rm{pure}}} +\lr{1-G^2} \frac{1}{2} +\var{S_p}.
\end{align}
The variances of $x_L^{\rm{pure}}$ and $p_L^{\rm{pure}}$ equal $1/\lr{2s}$ or $s/2$ depending on the phase of the squeezing, where $s=4$ in the experiment.
By inserting these experimental parameters and the variances of the final atomic state, we can calculate the added noise. The results are given in Table \ref{table2} for the two cases where we stored squeezed vacuum.

\begin{table}[h]
	\centering
		\begin{tabular}{|c|c|c|c|c|}
		\hline
& \multicolumn{2}{|c|}{Stored state} & \multicolumn{2}{|c|}{Added noise} \\ \hline

		$\phi$ & $\var{x_A^{\mathrm{fin}}}$ & $\var{p_A^{\mathrm{fin}}}$ & $\var{S_x}$ & $\var{S_p}$    \\ \hline
       0   &   2.02      &    0.52     & 0.08        &  0.29         \\ \hline
      90   &   0.72      &    1.90     & 0.13        &  0.32         \\ \hline
   	\end{tabular}
\caption{The table shows the measured final atomic  state variances $\var{x_A^{\mathrm{fin}}}$ and $\var{p_A^{\mathrm{fin}}}$ and the added extra noise  $\var{S_x}$ and $\var{S_p}$.} \label{table2}
\end{table}
The variances of $S_x$ and $S_p$ are quite small (less than one vacuum unit), and can be attributed to atomic decoherence and noise of the anti-squeezed light quadrature which can feed into the atomic $p_A^{\rm{fin}}$ due to imperfect electronic feedback.

\subsection{Local/non-local operators and cos/sine vs sideband picture}
Below we will describe the quantum memory protocol in the language of local atomic operators and the upper and lower sideband modes instead of using the non-local atomic operators and the cosine and sine light modes.
The local atomic operators and the non-local operators are connected by
\begin{align}
x_{A1} = & \frac{1}{\sqrt{2}} \lr{ x_{Ac} + p_{As} }, &  p_{A1} =  & \frac{1}{\sqrt{2}} \lr{ p_{Ac} - x_{As} },   & \nonumber \\
x_{A2} = & \frac{1}{\sqrt{2}} \lr{ x_{Ac} - p_{As} }, &   p_{A2} = & \frac{1}{\sqrt{2}} \lr{ p_{Ac} + x_{As} }   ,& 	 \label{eq:12cossin}
\end{align}
where the subscripts 1 and 2 refers to the first and second atomic ensemble, respectively.
We can also find the relation between the upper and lower sideband modes of light and its cosine and sine mode
\begin{align}
x_{Lc} = & \frac{1}{\sqrt{2}} \lr{ x_+ + x_- },  &   p_{Lc} =  & \frac{1}{\sqrt{2}} \lr{ p_+ + p_- },    & \nonumber \\
x_{Ls} = & \frac{1}{\sqrt{2}} \lr{ p_- - p_+ },  &   p_{Ls} = & \frac{1}{\sqrt{2}}  \lr{ x_+ - x_- }.  & \label{eq:+-cossin}
\end{align}
We write the input-output equations for the quantum memory as
\begin{align}
x_{Ac}^{\mathrm{fin}} = &  G  p_{Lc}^{\rm{pure}}  + O_{x},  & p_{Ac}^{\mathrm{fin}} = & -G  x_{Lc}^{\rm{pure}}   + O_{p}, & \nonumber \\
x_{As}^{\mathrm{fin}} = &  G  p_{Ls}^{\rm{pure}}  + O_{x},  & p_{As}^{\mathrm{fin}} = & -G  x_{Ls}^{\rm{pure}}   + O_{p}, & \label{eq:io2}
\end{align}
with the definitions
\begin{align}
O_x = & \sqrt{1-\frac{1}{Z^2}}x_A +\sqrt{1-G^2} p_{L}^{\rm{vac}} +S_x  & \nonumber \\
O_p = & \sqrt{1-G^2} x_{L}^{\rm{vac}}+S_p. &
\end{align}
Using $S_x \approx 0.1$ and $S_p \approx 0.3$, we find $\var{O_x}=0.44$ and $\var{O_p}=0.60$.
Equation \eqref{eq:io2} can be re-written by inserting Eqs.\ \eqref{eq:12cossin} and \eqref{eq:+-cossin}, and we find
\begin{align}
x_{A1}^{\mathrm{fin}} = &   G p_{+}^{\rm{pure}}  + \frac{ O_{x}+O_{p} }{\sqrt{2}}, &
p_{A1}^{\mathrm{fin}} = &  -G x_{+}^{\rm{pure}}  + \frac{ O_{p}-O_{x} }{\sqrt{2}}, & \nonumber \\
x_{A2}^{\mathrm{fin}} = &   G p_{-}^{\rm{pure}}  + \frac{ O_{x}-O_{p} }{\sqrt{2}}, &
p_{A2}^{\mathrm{fin}} = &  -G x_{-}^{\rm{pure}}  + \frac{ O_{p}+O_{x} }{\sqrt{2}}. & \label{eq:io12}
\end{align}
We see that the upper sideband is stored in the first atomic ensemble and the lower sideband is stored in the second ensemble.

\subsection{Storage of one part of an EPR-entangled pair}
In the experiment, the sender prepares and sends a two-mode entangled state to the receiver, who then stores it in his quantum memory. Although the memory is a true quantum memory as proven by its ability to outperform the classical benchmark, the noise added in the storage process to both EPR modes leads to a separable state of the two memory cells.  Note that for input states with the squeezing direction $\phi=0$ displaced up to $3.8$ (Table \ref{table1}), we are very close to having two displaced entangled atomic ensembles after the storage, since the parameter $E$ describing EPR-entanglement
\begin{align}
E \equiv&   \var{x_{A1}^{\rm{fin}}-x_{A2}^{\rm{fin}}}+\var{p_{A1}^{\rm{fin}}+p_{A2}^{\rm{fin}}} \nonumber \\
  =&     2\cdot  \left[\var{p_{Ac}^{\rm{fin}}}+\var{p_{As}^{\rm{fin}}} \right]           \nonumber \\
  =& 2\cdot\lr{0.52+0.52}=2.08,
\end{align}
is only slightly above 2.\\

Since outperforming the classical benchmark is sufficient to prove that the memory is capable of storing entanglement, we should be able to think of a modification of the experiment which can do exactly that. One example of such an experiment is the protocol, where Alice sends only one mode of the EPR-entangled pair to Bob for storage, and the other mode is sent to a third person, Charlie. In this case only one of the two entangled modes gets distorted by the memory. After the storage, Bob and Charlie perform a joint measurement to test whether there is entanglement between Bob's stored atomic state and Charlie's light state. In practice, Alice would have to separate the upper and lower sidebands, which could for instance be done using a cavity which transmits one sideband and reflects the other.
The initial entanglement of the upper and lower sidebands is characterized by
\begin{align}
E \equiv    &  \var{x_+^{\rm{pure}} - x_-^{\rm{pure}}} +\var{p_+^{\rm{pure}} -p_-^{\rm{pure}} } =2/s.
\end{align}
Alice then sends the EPR-entangled upper sideband  (together with a lower sideband in the vacuum state) to Bob who stores the upper sideband in the first ensemble (and the vacuum in the lower sideband in the second ensemble).
After the storage Bob and Charlie share the entanglement, since
\begin{align}
E\equiv    &  \var{x_-^{\rm{pure}} + p_{A1}^{\rm{fin}}} +\var{p_-^{\rm{pure}} -x_{A1}^{\rm{fin}}} & \nonumber \\
 = &   \frac{1}{2s} \lr{1+G}^2   + \frac{s}{2} \lr{1-G}^2     +   \var{O_{x}} +
 \nonumber \\
  +   &   \var{O_{p}}=1.51<2,
\end{align}
which has been calculated using the experimentally obtained parameters of our memory from Table \ref{table2}, Eq.\ \eqref{eq:io12}  and utilizing the fact that $\var{ x_+^{\rm{pure}} +x_-^{\rm{pure}}},\var{ p_+^{\rm{pure}} +p_-^{\rm{pure}}} = s$.
We conclude, based on the experimental performance of our quantum memory for the storage of both modes of the two mode entangled states, that if we instead stored only one part of the EPR-pair, one of the atomic memories would be entangled with the other part of the EPR-pair after the storage.

\subsection{Monotonicity of classical fidelity benchmark}

The following aims at proving that the classical benchmark is a non-increasing function of $d_{\max} $. That is, the broader is the a priori distribution in phase space, the more difficult it gets for a classical scheme to achieve a certain fidelity.
Here "classical scheme" refers either to entanglement breaking quantum channels or to channels whose Choi matrix has a positive partial transpose.

\begin{figure}
\centering
\includegraphics[scale=0.5]{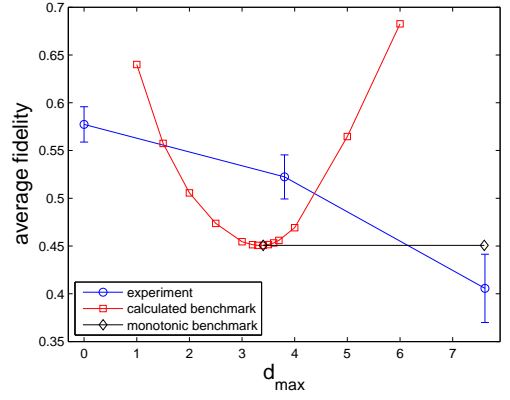}
\caption{\textbf{Fidelities.}
The fidelity $F$ calculated from experimental results is shown as circles connected by lines. The calculated benchmark values are shown as squares and the monothonic benchmark values are shown as diamonds.
The horizontal axis is the size $d_{\rm{max}}$ of the input distribution. Note that one vacuum unit of displacement corresponds to $d_{\rm{max}}=1/\sqrt{2}$.}
\label{fig:fidelities2}
\end{figure}

This supplement justifies the horizontal asymptote for the benchmark in Fig. 2 of the main text. Without this a rigorous treatment of the truncation errors (as in ref. Owari et al in the main text) would for large $d_{\max} $ lead to an increasing bound for the classical benchmark (see Fig.\ \ref{fig:fidelities2}).

Recall that our benchmark is based on the average fidelity
\begin{equation}
{F}(T):=\int_{\mathbb{R}^2} \hspace{-4pt}d\xi q(\xi)\int_{[0,2\pi)}\frac{d\theta}{2\pi}\;\tr{T\big({\cal N}_\lambda(\rho_{\theta,\xi})\big)\rho_{\theta,\xi}},
\end{equation}
where $\rho_{\theta,\xi}$ are the initial pure squeezed coherent states  with mean $\xi$ and orientation $\theta$, ${\cal N}_\lambda$ is an attenuation channel which decreases the intensity by a factor $\lambda$, $T$ is the channel corresponding to a hypothetical classical memory and $q(\xi)$ is the probability density distribution of the input alphabet in phase space.
The benchmark is then given by $$F:=\sup_T F(T), $$ where the supremum runs over all $T$'s corresponding to classically possible schemes.

We want to compare different values of $F$ corresponding to different distributions $q$, so we regard $F$ as a function of $q$ and write $F_q$.  Let us define
\begin{equation}
T'(\rho):=\int_{\mathbb{R}^2} d\eta\; p(\eta) W_\eta^\dagger T\big(W_{\sqrt{\lambda}\eta}\rho W_{\sqrt{\lambda}\eta}^\dagger\big)W_\eta,
\end{equation}
where $W_\eta$ is the Weyl operator which displaces by $\eta$ in phase space and $p(\eta)$ is some probability density distribution.
Note that $T'$ characterizes a classical, albeit coarse-grained, scheme if $T$ does. In this case
\begin{eqnarray}
F_q&\geq& \int_{\mathbb{R}^2} \hspace{-4pt}d\xi \; q(\xi)\int_{[0,2\pi)}\frac{d\theta}{2\pi}\;\tr{T'\big({\cal N}_\lambda(\rho_{\theta,\xi})\big)\rho_{\theta,\xi}}\\
&=&  \int\frac{d\theta}{2\pi}d\xi\; d\eta\; p(\eta)q(\xi) \tr{T\big({\cal N}_\lambda(\rho_{\theta,\xi+\eta})\big)\rho_{\theta,\xi+\eta}}\nonumber\\
&=& \int_{\mathbb{R}^2} \hspace{-4pt}d\xi \; q'(\xi)\int_{[0,2\pi)}\frac{d\theta}{2\pi}\;\tr{T\big({\cal N}_\lambda(\rho_{\theta,\xi})\big)\rho_{\theta,\xi}}\\ &=& F_{q'}(T),
\end{eqnarray}
where we used that $W_{\sqrt{\lambda}\eta}{\cal N}_\lambda (\rho)W_{\sqrt{\lambda}\eta}^\dagger={\cal{N}}_\lambda\big(W_{\eta}\rho W_{\eta}^\dagger\big)$ and we introduced the convolution
$$q'(\xi):=\int_{\mathbb{R}^2}d\eta\;q(\xi-\eta)p(\eta).$$
Choosing $T$ to be the optimal classical scheme for $q'$ we obtain that \begin{equation}
F_q\geq F_{q'},\label{eq:Fq}
\end{equation}
under the assumption that $q'$ is obtained from $q$ by convolution with another probability density $p$. If $q$ and $q'$, for instance, were two Gaussians, then Eq.\ (\ref{eq:Fq}) holds whenever $q'$ is broader than $q$ since $p$ can then be chosen to be a Gaussian whose variance is the difference between those of $q'$ and $q$.

Similar holds if, as in the experimental setup, $q$ is a flat-top distributions on a square $(-d,d]\times(-d,d]$. Taking the convolution with a discrete distribution $p(\eta)=\frac14 \sum_{i=1}^4\delta(\eta-\eta_i)$, where the $\eta_i$'s are the four corners of the square $[-d,d]\times[-d,d]$, leads to a flat-top distribution $q'$ on the square $(-2d,2d]\times(-2d,2d]$. This justifies to use the benchmark computed for $d_{\max}=3.8$ again for $d_{\max}=7.6$ as done in Fig. 2 of the main text.

\end{document}